\documentclass[12pt, draftclsnofoot, onecolumn]{IEEEtran}

\def\ifemptyarg[#1][#2]{%
	\if\relax\detokenize{#1}\relax
		#1
	\else
		#2
	\fi}
	
\def\ConstantComparison[#1]{
        \begin{subfigure}[t]{0.3\textwidth}
                \centering
                \includegraphics[width=\textwidth]{#1_constant_comparison}
                \caption{Comparison between the predicted values and the 
observed ones.}
                \label{fig:#1ConstantComparison}
        \end{subfigure} 
}

\def\ArOrderThreeComparison[#1]{
        \begin{subfigure}[t]{0.43\textwidth}
                \centering
                \includegraphics[width=\textwidth]{#1_ar_order_3_comparison}
                \caption{Comparison between the predicted values and the 
observed ones.}
                \label{fig:#1ArOrderThreeComparison}
        \end{subfigure} 
}

\def\MaOrderThreeComparison[#1]{
        \begin{subfigure}[t]{0.43\textwidth}
                \centering
                \includegraphics[width=\textwidth]{#1_ma_order_3_comparison}
                \caption{Comparison between the predicted values and the 
observed ones.}
                \label{fig:#1MaOrderThreeComparison}
        \end{subfigure} 
}

\def\ArmaOrderThreeThreeComparison[#1]{
        \begin{subfigure}[t]{0.3\textwidth}
                \centering
\includegraphics[width=\textwidth]{#1_arma_order_3_3_comparison}
                \caption{Comparison between the predicted values and the 
observed ones.}
                \label{fig:#1ArmaOrderThreeThreeComparison}
        \end{subfigure}
}

\def\ArmaGarchComparison[#1]{
        \begin{subfigure}[t]{0.3\textwidth}
                \centering
\includegraphics[width=\textwidth]{#1_arma_garch_comparison}
                \caption{Comparison between the predicted values and the 
observed ones.}
                \label{fig:#1ArmaArmaGarchComparison}
        \end{subfigure}
}

\def\ExponentialComparison[#1]{
        \begin{subfigure}[t]{0.3\textwidth}
                \centering
                \includegraphics[width=\textwidth]{#1_exponential_comparison}
                \caption{Comparison between the predicted values and the 
observed ones.}
                \label{fig:#1ExponentialComparison}
        \end{subfigure}
}

\def\ConstantHistogram[#1]{
        \begin{subfigure}[t]{0.3\textwidth}
                \centering
                \includegraphics[width=\textwidth]{#1_constant_histogram}
                \caption{Sizes of the errors of the predicted values.}
                \label{fig:#1ConstantHistogram}
        \end{subfigure}
}

\def\ArOrderThreeThreeHistogram[#1]{
        \begin{subfigure}[t]{0.3\textwidth}
                \centering
                \includegraphics[width=\textwidth]{#1_ar_order_3_histogram}
                \caption{Sizes of the errors of the predicted values.}
                \label{fig:#1ArOrderThreeThreeHistogram}
        \end{subfigure}
}

\def\MaOrderThreeHistogram[#1]{
        \begin{subfigure}[t]{0.3\textwidth}
                \centering
                \includegraphics[width=\textwidth]{#1_ma_order_3_histogram}
                \caption{Sizes of the errors of the predicted values.}
                \label{fig:#1MaOrderThreeThreeHistogram}
        \end{subfigure}
}

\def\ArmaOrderThreeThreeHistogram[#1]{
        \begin{subfigure}[t]{0.3\textwidth}
                \centering
                \includegraphics[width=\textwidth]{#1_arma_order_3_3_histogram}
                \caption{Sizes of the errors of the predicted values.}
                \label{fig:#1ArmaOrderThreeThreeHistogram}
        \end{subfigure}
}

\def\ArmaGarchHistogram[#1]{
        \begin{subfigure}[t]{0.3\textwidth}
                \centering
                \includegraphics[width=\textwidth]{#1_arma_garch_histogram}
                \caption{Sizes of the errors of the predicted values.}
                \label{fig:#1ArmaGarchHistogram}
        \end{subfigure}
}

\def\ExponentialHistogram[#1]{
        \begin{subfigure}[t]{0.3\textwidth}
                \centering
                \includegraphics[width=\textwidth]{#1_exponential_histogram}
                \caption{Sizes of the errors of the predicted values.}
                \label{fig:#1ExponentialHistogram}
        \end{subfigure}
}

\def\CleanCombined[#1][#2]{
        \begin{subfigure}[t]{0.43\textwidth}
                \centering
                \includegraphics[width=\textwidth]{#1_periods}
                \caption{#2}
                \label{fig:#1Periods}
        \end{subfigure}
}

\def\ConstantCombined[#1][#2]{
	\begin{subfigure}[t]{0.43\textwidth}
                \centering
                \includegraphics[width=\textwidth]{#1_constant_combined}
                \caption{#2}
                \label{fig:#1ConstantCombined}
        \end{subfigure}%
}

\def\ArOrderThreeCombined[#1][#2]{
	\begin{subfigure}[t]{0.43\textwidth}
                \centering
                \includegraphics[width=\textwidth]{#1_ar_order_3_combined}
                \caption{#2}
                \label{fig:#1ArOrderThreeThreeCombined}
        \end{subfigure}%
}

\def\MaOrderThreeCombined[#1][#2]{
	\begin{subfigure}[t]{0.43\textwidth}
                \centering
                \includegraphics[width=\textwidth]{#1_ma_order_3_combined}
                \caption{#2}
                \label{fig:#1MaOrderThreeThreeCombined}
        \end{subfigure}%
}

\def\ArmaOrderThreeThreeCombined[#1][#2]{
        \begin{subfigure}[t]{0.43\textwidth}
                \centering
                \includegraphics[width=\textwidth]{#1_arma_order_3_0_3_combined}
                \caption{#2}
                \label{fig:#1ArmaOrderThreeThreeCombined}
        \end{subfigure}
}

\def\ArmaGarchCombined[#1]{
        \begin{subfigure}[t]{0.43\textwidth}
                \centering
                \includegraphics[width=\textwidth]{#1_arma_garch_combined}
                \caption{Actual observations interleaved with predictions.}
                \label{fig:#1ArmaGarchCombined}
        \end{subfigure}
}

\def\ExponentialCombined[#1][#2]{
        \begin{subfigure}[t]{0.43\textwidth}
                \centering
                \includegraphics[width=\textwidth]{#1_exponential_combined}
                \caption{#2}
                \label{fig:#1ExponentialCombined}
        \end{subfigure}
}

\def\ExponentialSimpleCombined[#1][#2]{
        \begin{subfigure}[t]{0.43\textwidth}
                \centering
                
\includegraphics[width=\textwidth]{#1_exponential_simple_combined}
                \caption{#2}
                \label{fig:#1ExponentialSimpleCombined}
        \end{subfigure}
}

\def\ExponentialHoltsLinearCombined[#1][#2]{
        \begin{subfigure}[t]{0.43\textwidth}
                \centering
                
\includegraphics[width=\textwidth]{#1_exponential_holts_linear_combined}
                \caption{#2}
                \label{fig:#1ExponentialHoltsLinearCombined}
        \end{subfigure}
}

\def\SMCombined[#1][#2]{
        \begin{subfigure}[t]{0.43\textwidth}
                \centering
                \includegraphics[width=\textwidth]{#1_sm_combined}
                \caption{#2}
                \label{fig:#1SMCombined}
        \end{subfigure}
}

\def\MMCombined[#1][#2]{
        \begin{subfigure}[t]{0.43\textwidth}
                \centering
                \includegraphics[width=\textwidth]{#1_mm_combined}
                \caption{#2}
                \label{fig:#1MMCombined}
        \end{subfigure}
}

\def\MVNCombined[#1][#2][#3]{
        \begin{subfigure}[t]{0.43\textwidth}
                \centering
                
\includegraphics[width=\textwidth]{#1_#2_#3_mvn_combined}
                \caption{Actual observations and predictions based on them.}
                \label{fig:#1#2#3MVNCombined}
        \end{subfigure}
}

\def\MVNConditional[#1][#2][#3]{
        \begin{subfigure}[t]{0.43\textwidth}
                \centering
                
\includegraphics[width=\textwidth]{#1_#2_#3_mvn_conditional}
                \caption{Conditional bivariate normal distribution.}
                \label{fig:#1#2#3MVNConditional}
        \end{subfigure}
}

\def\KernelCombined[#1][#2][#3][#4]{
        \begin{subfigure}[t]{0.43\textwidth}
                \centering
                
\includegraphics[width=\textwidth]{#1_#2_#3_kernel_combined}
                \caption{#4}
                \label{fig:#1#2#3KernelCombined}
        \end{subfigure}
}

\def\NNCombined[#1][#2][#3][#4][#5]{
        \begin{subfigure}[t]{#4\textwidth}
                \centering
                
\includegraphics[width=\textwidth]{#1_#2_#3_nnet_combined}
                \caption{#5}
                \label{fig:#1#2#3ANNCombined}
        \end{subfigure}
}

\def\LinearCombined[#1][#2]{
        \begin{subfigure}[t]{0.43\textwidth}
                \centering
                
\includegraphics[width=\textwidth]{#1_linear_combined}
                \caption{#2}
                \label{fig:#1LinearCombined}
        \end{subfigure}
}

\def\Linear3D[#1][#2][#3][#4]{
        \begin{subfigure}[t]{0.43\textwidth}
                \centering
                
\includegraphics[width=\textwidth]{#1_#2_#3_linear_3d}
                \caption{#4}
                \label{fig:#1#2#3Linear3D}
        \end{subfigure}
}

\def\LinearAlternative3D[#1][#2][#3][#4]{
        \begin{subfigure}[t]{0.43\textwidth}
                \centering
                
\includegraphics[width=\textwidth]{#1_#2_#3_linear_alternative_3d}
                \caption{#4}
                \label{fig:#1#2#3AlternativeLinear3D}
        \end{subfigure}
}

\def\Kernel3D[#1][#2][#3][#4]{
        \begin{subfigure}[t]{0.43\textwidth}
                \centering
                
\includegraphics[width=\textwidth]{#1_#2_#3_kernel_3d}
                \caption{#4}
                \label{fig:#1#2#3Kernel3D}
        \end{subfigure}
}

\def\KernelAlternative3D[#1][#2][#3][#4]{
        \begin{subfigure}[t]{0.43\textwidth}
                \centering
                
\includegraphics[width=\textwidth]{#1_#2_#3_kernel_alternative_3d}
                \caption{#4}
                \label{fig:#1#2#3AlternativeKernel3D}
        \end{subfigure}
}

\newcommand{\Gateway}[1]{GW#1}

\newcommand{\Intel}{\emph{Intel}}
\newcommand{\Sensorscope}{\emph{Sensorscope}}
\newcommand{\Ball}{\emph{Ball}}
\newcommand{\Running}[1]{\emph{Running}#1}

\newcommand{\Constant}[1]{\emph{Constant}}
\newcommand{\LinearMethod}[1]{\emph{Linear}}

\usepackage{amsmath}
\usepackage{mathtools}
\usepackage{dsfont}
\usepackage{paralist}
\usepackage[table]{xcolor}
\usepackage{footnote}
\usepackage{array}
\usepackage{ragged2e}
\usepackage{bm}
\usepackage{textcomp}
\usepackage{hyperref}
\usepackage{subcaption}
\usepackage[pdftex]{graphicx}

\usepackage{afterpage}
\usepackage{rotating}
\usepackage{multirow}
\usepackage{lscape}
\usepackage{verbatim}

\usepackage{tikz}

\graphicspath{{./}{../figures/}{../plots/}}

\title{On the importance and feasibility of \\ forecasting data in sensors}
\author{Gabriel Martins Dias}

\author{\IEEEauthorblockN{Gabriel Martins Dias, Boris Bellalta and Simon 
Oechsner}
\IEEEauthorblockA{~\\Department of Information and Communication Technologies\\ 
Universitat Pompeu Fabra, Barcelona, Spain\\
Email: \{gabriel.martins, boris.bellalta, simon.oechsner\}@upf.edu}
}

\begin{document}

\maketitle

\begin{abstract}
The first generation of wireless sensor nodes have constrained energy 
resources and computational power, which discourages applications to process 
any task other than measuring and transmitting towards a central server.
However, nowadays, sensor networks tend to be incorporated into the Internet of 
Things and the hardware evolution may change the old strategy of avoiding data 
computation in the sensor nodes.
In this paper, we show the importance of reducing the number of transmissions 
in sensor networks and present the use of forecasting methods as a way of doing 
it.
Experiments using real sensor data show that state-of-the-art forecasting 
methods can be successfully implemented in the sensor nodes to keep the quality 
of their measurements and reduce up to $30\%$ of their transmissions, lowering 
the channel utilization. 
We conclude that there is an old paradigm that is no longer the most beneficial, 
which is the strategy of always transmitting a measurement when it differs by 
more than a threshold from the last one transmitted.
Adopting more complex forecasting methods in the sensor nodes is the 
alternative to significantly reduce the number of transmissions without 
compromising the quality of their measurements, and therefore support the 
exponential growth of the Internet of Things. 

%
\end{abstract}

\section{Introduction}

Big Data analytics are beginning to be applied as part of the process of the 
sensor nodes analysis and management, for example, to increase the number of 
measurements when the environment is changing.
The efficiency of such data analytic methods is highly correlated with the 
quality of the data used, i.e., reported by the sensors~\cite{Hazen2014}.
Among other aspects, the information quality depends on the temporal relevance, 
the data resolution and the chronology of the data~\cite{Kenett2014}.

At the same time, sensor nodes have evolved in the last years from devices with 
constrained energy and memory resources~\cite{XBowTelosB}, to the
point where some modern hardware can harvest energy and work autonomously 
for longer periods~\cite{waspmote}.
According to~\cite{Rault2014}, the most recent technologies of wireless power 
charging should allow the energy constraint of the sensor nodes to be overcome 
in the future.
Moreover, besides the hardware evolution, the incorporation of Wireless Sensor 
Networks (WSNs) into the Internet of Things (IoT)~\cite{Bellavista2013} 
accelerates the evolution of the sensor networks, since smartphones and 
household appliances can easily become sensor nodes, by generating measurements 
and communicating them to their neighbors.

Although the hardware limitations tend to disappear, the medium access has been 
named as one of the key challenges in the next generations of 
wireless networks due to the increase on the number of wireless devices and 
traffic profiles~\cite{Bellalta2015}.
Hence, the access to the data produced by neighboring wireless sensor nodes 
might remain an issue in the next generation of sensor networks.
Also due to the limited channel resources, sensor nodes are usually programmed to 
transmit their measurements as rarely as possible.
For example, the most intuitive solution is to transmit the measurements only 
when they differ by more than a certain threshold from the last value observed, 
i.e., when the data probably contains new--and valuable--information.

Driven by the sensor nodes' evolution, there are several 
approaches to reduce the number of transmissions by adopting complex 
forecasting methods.
They substitute the simpler strategy of avoiding a transmission if the current 
measurement is the same--or very similar--to the last one 
transmitted~\cite{Li2013,Liu2005,Askari2011,Li2009}.
As detailed in our previous work~\cite{Dias2015}, theoretically, high accuracy 
forecasts can reduce the amount of transmissions by around $30\%$ in an ordinary 
scenario and by up to $85\%$ in sporadic cases.

Meanwhile, there are no comparison studies that show whether such a high 
accuracy can be achieved in practice or not.
In this work, we show the potential benefits of adopting forecasting methods in 
different scenarios through experiments using real sensor data.
To do that, we first illustrate the importance of analyzing the effectiveness 
of the forecasts when applied to real use cases, i.e., based on the numerical 
measures of forecast accuracy, we adopt a strategy to evaluate the practical 
benefits of the forecasting methods to sensor networks that can be 
adopted as a reference in similar scenarios.
For instance, considering that the sensors' resolution is the smallest interval that 
can be reliably measured, their resolution is the highest measurement quality that 
a sensor network can provide and the smallest error accepted in the forecasts.
Finally, our main contribution is a broad study about how much sensor networks 
can benefit by forecasting measurements in the sensor nodes to diminish their 
number of transmissions without reducing the quality of their measurements.

Our results reinforce the idea that it is possible to shift part of the data 
computation to the sensor nodes, reducing the number of transmissions and the 
channel utilization.
For the future on the Internet of Things, this work represents a step towards a 
distributed solution that can help to detain the significant increase in the number 
of transmissions and consequential quality reduction.



In Section~\ref{sec:background}, we clarify which kind of application we 
expect the sensor networks to be used for and describe the datasets that will 
be evaluated further.
After that, we explain what forecasts are and their difference to ordinary 
predictions, as well as introduce the terms and methods currently adopted in 
data science applications in Section~\ref{sec:forecast-methods}, before 
detailing how important their use is in sensor networks in 
Section~\ref{sec:importance-of-forecasting}, and describing the related work in 
Section~\ref{sec:related-work}.
Then, using real data collected in other studies, we synthesize the evolution 
of the sensor nodes based on their memory and computing capabilities, and 
observe how it can impact the results obtained by the state-of-the-art 
forecasting methods that are broadly used in other applications.
Based on the accuracy of the results shown in Section~\ref{sec:results}, we 
study their effectiveness in sensor network applications in 
Section~\ref{sec:feasibility}.
Finally, we draw conclusions about their adoption according to the trade-off 
between their computational complexity and their potential accuracy in 
Section~\ref{sec:conclusion}.

\section{Sensor network applications}
\label{sec:background}

We use the term sensor network to denote any set of sensor nodes that can 
measure environmental parameters and report them (through their neighbors, if 
necessary) to a gateway~(\Gateway{}) that works as a central server, collecting 
and storing all the data.
Such networks, however, may have different applications, measure different data 
types and be influenced by contrasting environmental circumstances.
In order to categorize their characteristics and requirements, we classify 
sensor network applications into two broad classes according to their nature: 
monitoring and tracking.

As defined in~\cite{Yick2008}, monitoring applications comprise mainly ``indoor 
and outdoor environmental monitoring, health and wellness monitoring, power 
monitoring, inventory location monitoring, factory and process automation, and 
seismic and structural monitoring''. 
Hence, in such applications, it is more common to encounter temperature, 
relative humidity, light, solar radiation, wind speed and soil 
moisture sensors, among others, that can measure environmental parameters;
and the data types are usually periodic, i.e., each type follows a similar 
pattern through the days (or weeks) that could be used, for example, to change 
their sampling rate at a certain time of the day, if the sensor nodes had 
enough computational power to store enough data and compute such a decision.

On the other hand, tracking applications include especially human tracking, 
battlefield observation (e.g., enemy tracking), animal tracking and car tracking 
in smart cities.
This kind of application usually requires more powerful sensors, such as 
cameras, microphones and radio-frequency identification, and is less tolerant to 
delays and single point of failures.
In many target tracking applications, the data is sensed by only one sensor 
node at a time, differently from monitoring applications that use several 
sensors to measure the environmental parameters.
As a result, the computation in the sensor nodes is heavier, because they tend 
to oversample the data and avoid missing variations that will eventually happen.
Moreover, the data is usually processed and only the relevant information is 
transmitted to the \Gateway{s}~\cite{Bhatti2009}.


Both application types have in common the data heterogeneity in terms of 
scales, since their values may be stored as nominals, ordinals, intervals 
or ratios~\cite{Kirch2008}.
Moreover, different datasets may use different units of measurement, for 
example, distances could be represented in kilometers or miles, temperatures 
could be represented in Kelvin, Fahrenheit or Celsius, and time intervals might 
be represented in seconds, minutes, hours or days.
In conclusion, the difference between sensor measurements does not stem only 
from the origin of the data, but also from its representation.

We highlight a special use of sensor networks that may be incorporated 
in both application types: the event detection.
Events can be detected by the \Gateway{} after analyzing the data collected 
by the sensors, or by the sensor nodes, if they compute the data locally.
The detection of an event is tightly tied to the scenario where it is applied 
and requires deep domain knowledge, including its causes and consequences.
Missing or wrongly reporting an event (i.e., false negatives and false 
positives) have inherent costs that impact the operation of the system.
Given that our goal is to find a solution that would satisfy as many 
applications as possible, we do not focus on the act of detecting an event, 
because it would imply on assessing different (higher) costs generated by false 
positives and false negatives.
We aim our attention at the data collection for general purposes, presuming 
that it can be adopted in a sensor network that detects events, if needed.

To represent monitoring applications, we will adopt two different datasets 
obtained from real world deployments with WSNs: the \Intel{} and the 
\Sensorscope{} datasets.
To represent the tracking applications, we will use the \Ball{} 
dataset, which was synthetically created based on a model of projectile 
launching, and the \Running{} dataset, which was collected using a GPS 
monitor carried by a person while running through a city.

\subsection{Intel data}

The first dataset was extracted from the experiments described 
in~\cite{IntelLabData:2004:Misc} and encompasses the temperature, relative 
humidity and luminance collected by sensor nodes inside an office during 
consecutive days.
The whole dataset contains around $2.3$ million readings done during $37$ 
consecutive days by $54$ sensor nodes that transmitted their measurements every 
$30$ seconds and has been broadly used in several works in the field
~\cite{Askari2011,Yann-Ael2005,Deshpande2004,Jiang2011,Min2010}.

In this work, only the temperature values will be used. 
We selected five consecutive days and observed the data collected by 
three nodes to illustrate the performance of the predictions. 
Two nodes were selected according to the variance in their temperature 
measurements, i.e., those with the lowest and the greatest variance, and the 
third one was randomly picked (respectively, sensor nodes $35$, $21$ and $40$).
The missing values were linearly interpolated and summed to a small white noise.
In Figure~\ref{fig:intel_temperature}, we can observe the data collected by 
the sensors vary between $15^{o}$C and $37^{o}$C.
Even though the measurements do not look very similar, it is possible to see 
that there is a daily pattern in which their values and variances increase 
during the day and are more stable and similar at the beginning and at the end 
of the days.

\subsection{Sensorscope data}

The Sensorscope data was collected by wireless sensor nodes in a deployment 
made on a rock glacier in Switzerland~\cite{Barrenetxea2008}.
The WSN was composed by $10$ sensor nodes specially designed for environment 
monitoring.
The experiment lasted $5$ days and each sensor node reported its measurements 
every $2$ minutes, which resulted in over $3000$ reports per node, each of 
them including $8$ different measurement types: temperature, solar radiation, 
relative humidity, soil moisture, watermark, rain level, wind speed and wind 
direction.

We used the temperature values of three nodes to illustrate the performance 
of the predictions.
The nodes selected were the ones that presented the lowest and the greatest 
variance in their measurements, and last one was randomly chosen (respectively, 
sensor nodes $5$, $7$ and $15$), similar to how we selected the nodes in the 
\Intel{} dataset.
Again, the missing values were linearly interpolated and summed to a small 
white noise.
Figure~\ref{fig:sensorscope_temperature} shows that, compared with the 
temperatures observed in the \Intel{} dataset, the values are much lower 
(between $-12^{o}$C and $12^{o}$C), which is explained by the sensors' 
localization and the nature of the experiments.
Moreover, there are less abrupt changes, although the presence of the sun 
clearly changes the values and increases their variance during the days.

\afterpage{
\begin{figure}[t]
        \centering
        \begin{subfigure}[t]{0.85\textwidth}
                \centering
                \includegraphics[width=\textwidth]{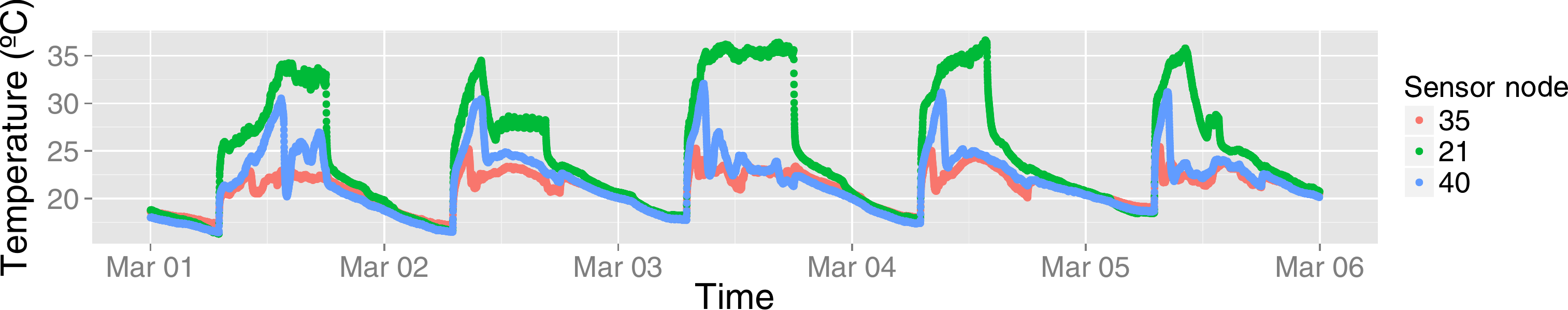}
                \caption{\Intel{}: Temperature measured by 
three sensors in an office~\cite{IntelLabData:2004:Misc}.}
                \label{fig:intel_temperature}
        \end{subfigure}%
        \qquad
        \begin{subfigure}[t]{0.85\textwidth}
                \centering
\includegraphics[width=\textwidth]{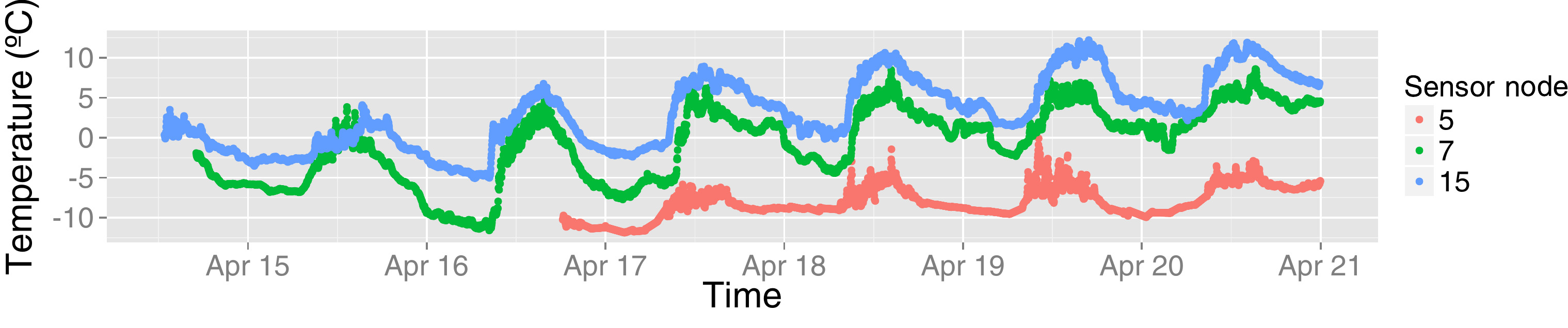}
                \caption{\Sensorscope{}: Temperature measured by three 
sensors in a mountain~\cite{Barrenetxea2008}.}
                \label{fig:sensorscope_temperature}
        \end{subfigure}%
	\qquad
	\begin{subfigure}[t]{0.75\textwidth}
			\includegraphics[width=\textwidth]{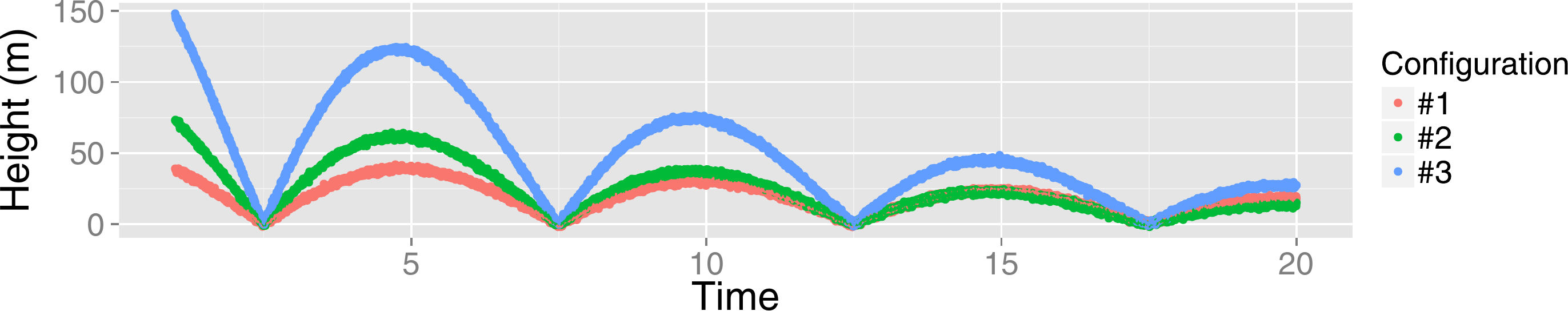}
			\caption{\Ball{}: Synthetic data created to imitate 
bouncing 
objects.}
			\label{fig:obj_data}
	\end{subfigure}%
	\qquad
	\begin{subfigure}[t]{0.75\textwidth}
			\includegraphics[width=\textwidth]{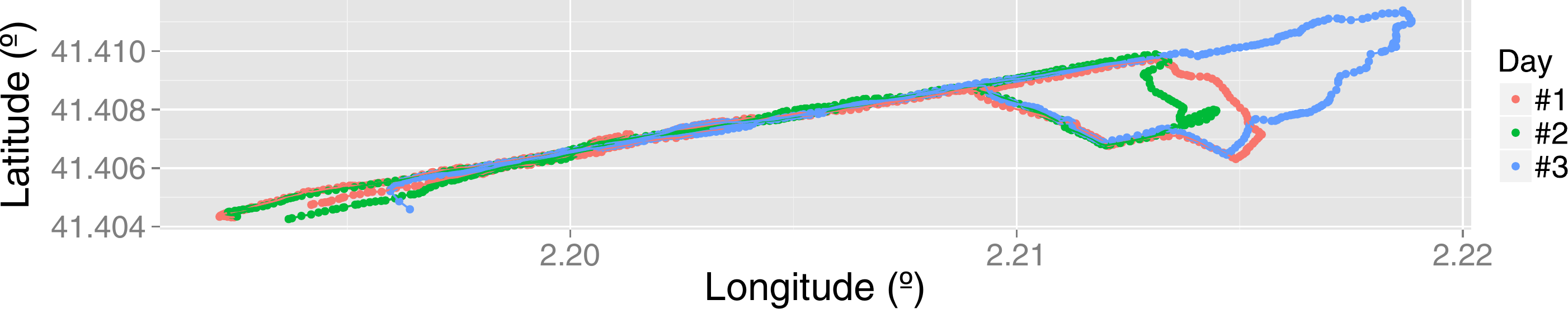}
			\caption{\Running{}: GPS coordinates of a person during 
street 
runs.}
			\label{fig:running}
	\end{subfigure}%
	\qquad
	\begin{subfigure}[t]{0.75\textwidth}
			\includegraphics[width=\textwidth]{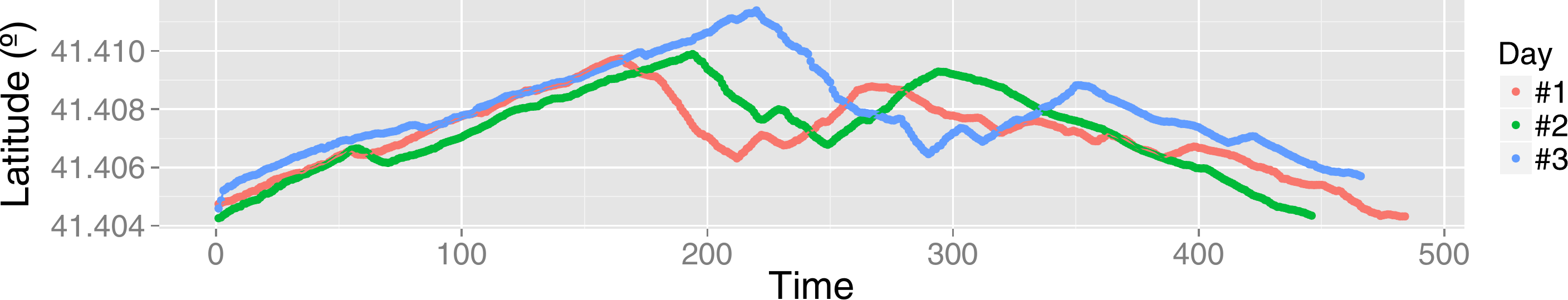}
			\caption{\Running{ (latitude)}: Geographic latitude of 
a 
person.}
			\label{fig:running-latitude}
	\end{subfigure}%
	\qquad
	\begin{subfigure}[t]{0.75\textwidth}
			\includegraphics[width=\textwidth]{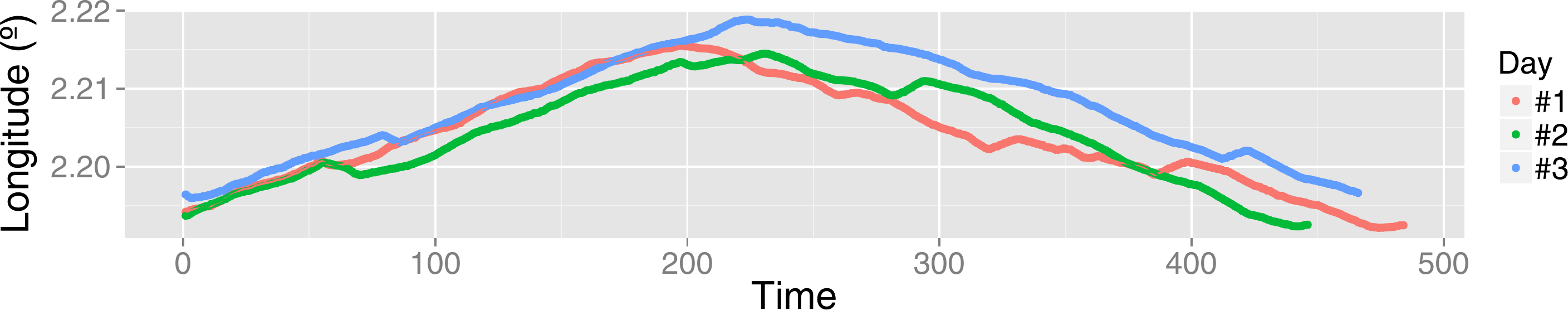}
			\caption{\Running{ (longitude)}: Geographic longitude 
of 
a person.}
			\label{fig:running-longitude}
	\end{subfigure}%
	\caption{Datasets used to illustrate the different applications.}
	\label{fig:datasets}
\end{figure}%
\clearpage
}

\subsection{Ball movement}

The first dataset used to represent tracking applications was synthetically 
created in order to simulate an object bouncing on the floor a few times. 
The data is intended to simulate an object being tracked and can be 
thought as the vertical position of a ball that hits the floor after being 
dropped from a certain height.
The data points were calculated using the formula of a pendulum with 
exponential decay~\cite{Baker2005}:

\begin{equation}
\theta(t) = \theta_0~\frac{| cos( 2\pi \lambda~t) |}{e^{\gamma~t}} + 
\varepsilon_t,
\end{equation}
where $\theta_0$ is the initial amplitude, $\lambda$ is the frequency, 
$\gamma$ represents the decay and $\varepsilon_t$ is an additive zero-mean, 
unit variance Gaussian white noise.
In order to reproduce different types of movements, we generated three 
sequences of data, each one with $2800$ data points.

Figure~\ref{fig:obj_data} shows the values in the \emph{Ball} dataset set with 
frequency $\lambda = 0.1$ hertz and sampled once every second.
The first set of points is based on a movement with initial amplitude $\theta_0 
= 50$ meters and decay $\gamma = 0.05$.
The second set of points has greater initial amplitude and decay ($\theta_0 = 
100$ meters, $\gamma = 0.1$), which means that the object moves faster, 
resulting on sparser data that may be less predictable.
Finally, the third set of points illustrates the fastest object, which has a
decay $\gamma = 0.1$ and the initial amplitude is $200$ meters, which means that 
the changes are more abrupt and less predictable than in the others.
Besides these differences, their periodicity can be clearly noticed in the 
plot.

\subsection{Street runner}

This dataset consists on the position of a person while running across the 
city of Barcelona.
The data was collected by a GPS device taken by a person in three different 
days and each observation was registered in an interval between $1$ and $5$ 
seconds after the last one, summing up to $480$ data points.
Even though the trajectories are similar, the measurements contain noise and 
variations that are expected to be encountered in other applications for object 
tracking.

In Figure~\ref{fig:running}, the \emph{Running} dataset is shown.
The different trajectories among the days are illustrated and it is possible to
observe their internal similarity and the absence of periodicity.
As we can observe in Figure~\ref{fig:running-latitude}, changes in the latitude 
are more abrupt and do not follow any pattern.
On the other hand, Figure~\ref{fig:running-longitude} shows that the longitude 
varies almost linearly in time and is more intuitive than the changes in the 
latitude.

\section{Forecasting methods - Background} 
\label{sec:forecast-methods}

The term \emph{prediction} can either refer to the process of inferring missing 
values in a dataset based on statistics or empirical probability, or to the 
estimation of future values based on the historical data.
The latter mechanism is also called \emph{forecast} and it is the class of 
predictions we will refer to in this work.
In summary, a forecast is a specific type of prediction in which the 
predicted values will be observed only in the future.
A forecast differs from the other predictions because, when 
estimating the future, a wider range of possible values must be considered, 
given the uncertainty about the factors that may impact the scenario under 
consideration.
In order to make it clear to the reader, in this Section we clarify a set of 
terms that will be often used in this work.

\subsection{Time series} 
A time series ($X$) is a sequence of data points, typically consisting of 
observations made over 
a time interval and ordered in time~\cite{Box2008}. 
Each observation is usually represented as $x_t$, where the observed value $x$ 
is indexed by the time $t$ at which it was made.

\subsection{Information criteria} 
Information criteria are measures used to estimate the information loss if the 
time series is modeled by a set of parameters $\theta$.
The estimated information loss is applied to infer the relative quality of the 
parameters and choose the best option given a set of candidates.
Assuming that the future data will have the same characteristics as the 
observations already made, the set of chosen parameters minimizes the 
information loss, which tends to improve the accuracy of the forecasts.
In some cases, the number of parameters is also taken into account, i.e., using
less parameters may be considered an advantage, because it avoids overfitting 
the training data. 

Examples of information criterion measures are the Akaike Information Criterion 
(AIC); the Bayesian Information Criterion (BIC); and the AIC with a correction 
for finite sample sizes (AICc)~\cite{Hyndman2008}.
In our experiments, we adopted the AICc as the information criterion to select 
the most proper set of parameters. 

\subsection{Forecasting method} 
A forecasting method ($F$) is a function that produces forecasts  based on two 
input values: a time series ($X$) and a set of parameters $(\theta)$.
The values of $\theta$ are usually chosen based on the evaluation provided by 
an information criterion measure, given $X$.

\subsection{Forecasting model} 
A forecasting model ($f$) is an instance of a forecasting method $F$, such that 
$f_{\theta}(X) = F(X, \theta)$. 
Every forecasting model is deterministic: its output depend only on the set 
of observed values.
Forecasting models (also called \emph{time series prediction 
models}~\cite{Box1990,Makridakis1998}) use time series as input to predict 
future values, which are represented as a function of the past observations and 
their respective time, i.e., $(\hat{x}_{t+1} \ldots \hat{x}_{t+i}) = 
f_{\theta}(x_{t}, \ldots, x_{t-k})$, where $\hat{x}_{t+1} \ldots \hat{x}_{t+i}$ 
are the forecasts for the period between $t+1$ and $t + i$.

\subsection{Accuracy} 
Accuracy measures are used to evaluate the quality of the predicted values 
based on the difference between the predicted value and the observed value, 
i.e., the error $e_t = x_t - \hat{x_t}$~\cite{hyndman2014forecasting}.
When necessary, the Mean Absolute Percentage Error (MAPE) will be used to 
compare the errors of different data types, because it is not scale-dependent.

\section{Importance of forecasting}
\label{sec:importance-of-forecasting}

\begin{figure}[t]
	\centering
	\includegraphics[width=0.21\textwidth]{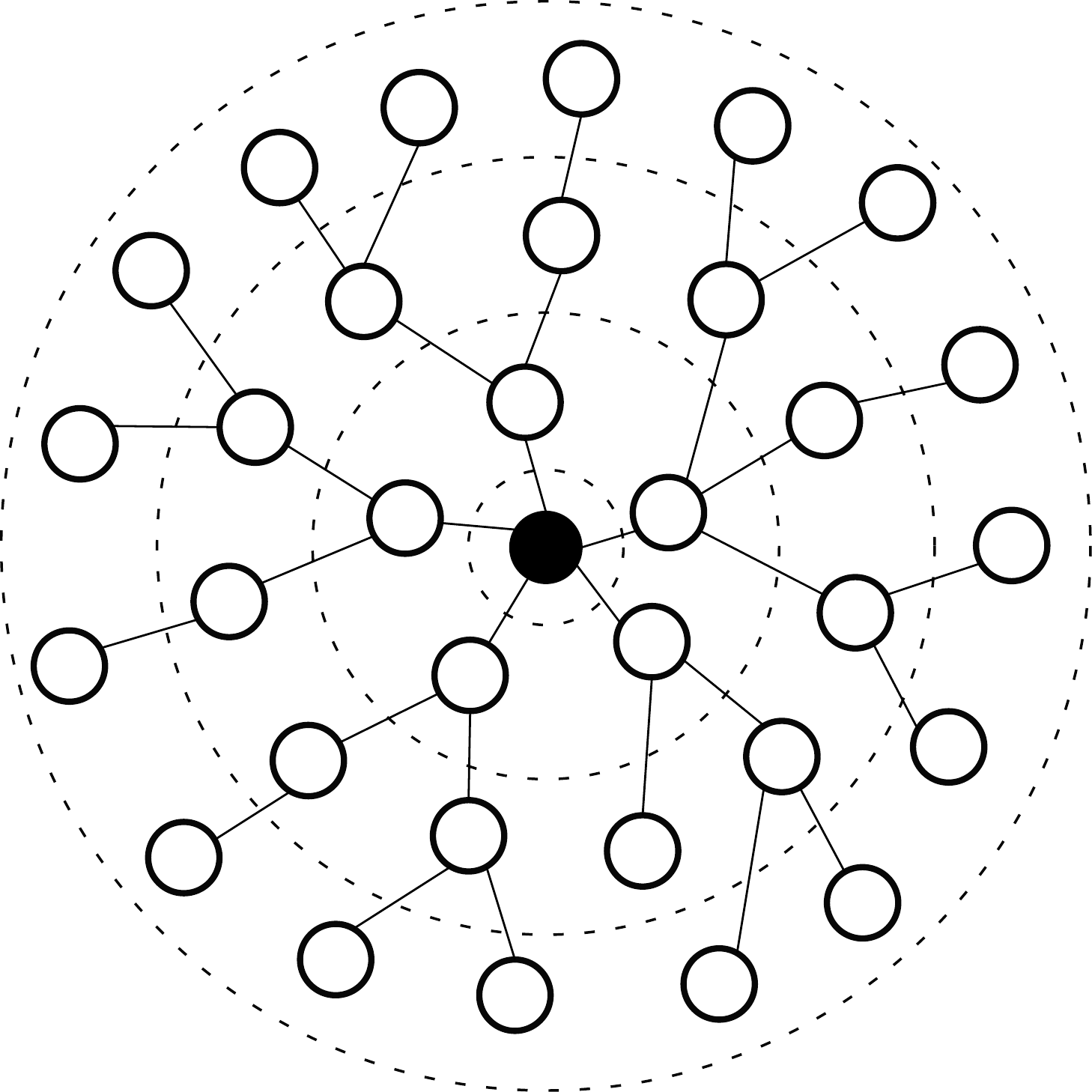}
	\caption{Sensor network model based on the density of sensor
nodes ($C$) and their coverage. The dark circle represents the \Gateway{} 
in a network with $3$ rings ($D = 3$, $C = 5$).}
	\label{fig:spanning-tree}
\end{figure}

In order to illustrate the potential growth in the number of transmissions, we 
will adopt the ring model for sensor network topologies presented 
in~\cite{Langendoen2010} and extended in~\cite{Dias2015}.
The ring model is based on the average number of neighbors ($C$) of a sensor 
node and on the number of hops from the \Gateway{} to the furthest nodes 
($D$), as illustrated in Figure~\ref{fig:spanning-tree}.
By definition, any transmission initiated by a sensor node in ring $d$ will 
generate, at least, $d - 1$ more transmissions to reach the \Gateway{}, 
resulting in, at least, $d$ transmissions in total.
Therefore, assuming that the sensor nodes are uniformly distributed in the 
plane and that there are $C + 1$ nodes in the unit disk, the first ring will 
contain $C$ nodes, and subsequently the number of nodes $N_d$ in ring $d$ can 
be calculated based on the surface area of the annulus\footnote{The region 
bounded by two concentric circles.}:

\begin{equation}
N_d =  
\begin{cases} 
0, & \text{if } d = 0\\ 
Cd^2-C(d-1)^2 = C(2d-1), & \text{otherwise.}
\end{cases}
\end{equation}
Hence, adding a new ring to a network with $D - 1$ rings and density 
$C$ represents $C(2D-1)$ new sensor nodes in a total of $CD^2$ nodes.

Recall that a transmission made by a sensor node in ring $d$ triggers other 
transmissions in the network, which sums up to, at least, $d$ new transmissions.
Considering that the sensors are programmed to transmit uniformly once 
every unit of time, all the sensors in the new ring $d$ together will trigger 
$Cd(2d-1)$ new transmissions.
Summing this expression from $d = 0$ to $d = D$ gives us that the total 
number of transmissions in the sensor network during a unit of time is

\begin{equation}
 \mathop{\sum_{d=0}^{D} Cd(2d-1)} = \frac{2}{3}~CD^3 - 
\frac{1}{2}CD^2\text{,}
\end{equation}
which finally shows that, considering a linear growth in the number of rings,  
the number of sensor nodes grows quadratically while the number of 
transmissions grows cubically.

In conclusion, the number of transmissions impacts directly the efficient use 
of spectrum resources, which (besides the energy consumption of the 
sensor nodes) is still one of the key challenges that affects 
the next generation of wireless networks, for instance, WLANs, 4G and 5G 
networks, as well as traditional multihop WSNs~\cite{Bellalta2015}.

Forecasting sensed data is a potential candidate to shorten such an 
increase in the number of transmissions, which is reinforced by its presence 
in real world sensor network applications. 
That is, most of the sensor networks forecast missing data, even though it is 
not explicitly addressed as an issue. 
It happens, for example, when a sensor measurement cannot be sampled on demand
and the value returned by the system to an ordinary query is the latest one (or 
a combination of some of them), which has been reported by the sensor nodes 
some seconds (or minutes) earlier.
In other words, it is the same technique used by the \Constant{} method: the 
system simply assumes that there was no change in the environment after the 
last observation~\cite{hyndman2014forecasting}. 
This behavior in turn is exploited by the sensors that avoid unnecessary 
transmissions and transmit only when a measurement differs by more than a fixed 
tolerance threshold.


The motivation for exploring advanced forecasting methods to diminish the 
number of transmissions is detailed in our previous work~\cite{Dias2015}: based 
on the ring model, accurate predictions can reduce the amount of transmissions 
by around $30\%$ in the average case and up to $85\%$ in exceptional cases.
The backbone of the model consists of an application of the central limit 
theorem and the law of large numbers, but no forecasting method was tested to 
observe whether the ideal accuracy could be achieved.
Therefore, having clarified that the use of predictions can benefit the 
sensor networks, we point out the unanswered question: ``Is it feasible to 
forecast the sensors' measurements in the sensors?''
Discarding the access to other data sources that could improve the forecasts 
accuracy, the list of forecasting methods is reduced and the results are 
occasionally worsened by the sensor nodes' constrained hardware.

For the next generation of sensor applications, a detailed study is necessary   
to show how accurate can be the forecasts made by the sensors and their 
respective cost-benefit analysis.
In the following, we will study the feasibility of using complex forecasting 
methods in the sensor nodes in order to reduce the number of transmissions 
without reducing the quality of their measurements.
Before that, in the next Section, we show how the recent works have adopted 
forecasting methods to reduce the data transmitted by their sensor nodes 
and how their operation could be improved.

\section{Related work}
\label{sec:related-work}

Forecasting methods have been adopted as part of data reduction techniques for
sensor networks.
In the beginning, the \Constant{} method was adopted in several mechanisms, 
because it corresponds to the constrained energy and power resources of the 
sensor nodes.
That is, the main goal of the first applications was to reduce the number of 
transmissions without making any complex computation in the sensor 
nodes~\cite{Lee2003,StrategiesXu2004,Jain2004a}

With the evolution of the sensor nodes, their energy and power constraints 
became less deterrent to the shift of some computation from the \Gateway{}.
As a consequence, several authors started to adopt and test enhanced methods to 
make more accurate predictions in the sensors.
For instance, the AutoRegressive Integrated Moving Average (ARIMA) is the 
forecasting method adopted in~\cite{Li2009}.
In that work, the \Gateway{} is responsible for generating the 
forecasting models before transmitting the parameter values to the sensor nodes.
After receiving the parameters, the sensor nodes are able to compute the same 
predictions as the \Gateway{} and locally check their accuracy. 
Finally, they only transmit the measurements if the forecast was
inaccurate.
Similarly, the mechanisms presented in~\cite{Li2013,Liu2005,Askari2011} use 
the traditional ARIMA as the method for predicting future measurements.
In~\cite{Askari2011}, the authors recognized the limitations of the traditional 
forecasting methods and suggested the adoption of an Artificial Neural Network 
(ANN) model when the predictions using ARIMA fail, which in turn requires more 
computational resources from the sensor nodes.

More recently, the mechanism presented in~\cite{Bogliolo2014} computes the 
forecasting models in the sensor nodes before transmitting them to the 
\Gateway{}.
That is, the sensor nodes became completely autonomous and independent of the 
data computation made in the \Gateway{}.
On the other hand, the choice of the method is still restricted by the sensor 
nodes' memory and processing power limitations, which also narrows the range of 
situations in which it can be successfully adopted.

Following the approaches of shifting the data computing to the sensor nodes, 
in~\cite{Aderohunmu2013}, a real WSN was deployed and part of its energy was 
saved through the reduction in the number of transmissions using different 
prediction methods: \Constant{}, weighted moving average, ARIMA and Exponential 
Smoothing (ES). 
The experimental results showed that the \Constant{} method was the best 
option in the use case considered by the authors, due to its better accuracy 
and the higher energy consumption of the others.
Even though there is a clear contribution towards the (non-)adoption of 
complex forecasting methods in sensor networks, the results are specific to the 
application presented by the authors and different use cases cannot be 
inferred from their study.

After observing the evolution of the sensor networks and the adoption
of more complex prediction methods in the field, we noticed that there was
a need to develop a broad study about the prediction algorithms from
a statistical point of view that could serve as a reference for future 
developments in the integration of data prediction methods into sensor networks.
Comparisons between the state-of-the-art algorithms have been also shown in the 
M3's competition results~\cite{Makridakis2000} and, although they show that the 
forecasts perform well in several cases, they do not consider the limited data 
access and processing power inherent to the sensor nodes.
In the next Section, we use sensor network data to present a comparison between 
the forecasting methods presented before under the constrained sensor nodes' 
perspective.

\section{Experimental results}
\label{sec:results}

The first step to validate the feasibility of using forecasting methods was to 
systematically apply the different methods over splits of the data collected 
by real sensors in different experiments.
Each split is defined by a \emph{history} plus a \emph{window}: the former 
represents the measurements made in the past and the latter the measurements to 
be forecast.
The idea of observing different \emph{history} and \emph{window} lengths is to 
synthesize the evolution of the sensor nodes' memory and computational 
capacities.
Hence, the experimental results illustrate the comparison between different 
configurations and how the sensor nodes' evolution impact the forecasts 
accuracy.

\subsection{Parameter study}

From now, we define as a \emph{scenario} each combination of \emph{history} and 
\emph{window} lengths observed in a \emph{dataset}.
In the experiments, for example, a scenario with history length of $100$ and 
window length equal to $10$ in the \Ball{} dataset has been represented by 
$200$ splits of data randomly picked from that dataset.
Before testing each scenario under different forecasting methods, we detail 
their characteristics in the following.

\subsubsection{Datasets}

The tests were made using all the datasets presented in 
Section~\ref{sec:background}: 
\emph{Intel}, \emph{Sensorscope}, \emph{Ball} and \emph{Running}.
Each dataset is composed by $3$ \emph{groups} of data, i.e., 
\emph{Intel} and \emph{Sensorscope} have $3$ different sensors each, 
\emph{Ball} was constructed using $3$ different parameter configurations and 
\emph{Running} contains data from $3$ different days, which was separated 
along two dimensions (latitude and longitude), and considered as two different 
sets of data.
The forecasts were made based on $200$ splits of data randomly picked from 
each group.
This setup represents the data heterogeneity expected in different sensor 
networks.

\subsubsection{Forecasting methods}

In our experiments, we tested the forecasting methods that are broadly used 
in data applications, thanks to their scalability and reliability: the 
\Constant{}, the \LinearMethod{}, the Simple Mean (SM), the Exponential 
Smoothing (ES), the AutoRegressive Integrated Moving Average (ARIMA) and the 
Artificial Neural Networks (ANNs) methods.
All of them are explained in detail in~\cite{hyndman2014forecasting}.

Different from the others, the ANNs are soft computing solutions that cannot be 
bounded by a computational time limit~\cite{Jain1996,Haykin1999}.
Therefore, besides it, all the computing complexities are summarized 
in Table~\ref{table:time-series-methods}.
Usually, the complexities are given in function of the number of values used to 
generate a forecasting model ($h$) and the number of values that will be 
forecast ($w$).
In some cases, other specific method parameters ($p$, $q$ and $k$) also impact 
the algorithms' complexity.
In short, their values can vary, but usually  $p \in [0,2]$,  $q \in [0,2]$ and 
$k \in [0,10]$.

\subsubsection{History}
The \emph{history} is the set of data points used during the learning phase 
and its length impacts the preprocessing time complexity, i.e., the computing 
time required to find the best prediction model or to set up the model 
parameters.
Thus, the simplest methods, such as the \Constant{} and the \LinearMethod{}, 
are not affected by the history length, but the time spent to set up an ARIMA 
model increases quadratically according to the history length and must be 
considered before its adoption in the simplest sensor nodes.
We considered cases in which the history length varied among $5$, $10$, $20$, 
$50$, $100$, $200$, $500$ and $1000$.


\subsubsection{Window}
The \emph{window} is the set of values that must be forecast and represent 
future measurements.
Hence, they are not considered at the moment in which the forecasts are 
produced, but only to measure their accuracy.
We experimented scenarios where $1$, $5$, $10$, $20$, $50$, $100$, $200$, $500$ 
and $1000$ values were predicted at a time.
The window length might affect the runtime, which can be either constant or 
increase linearly, as shown in Table~\ref{table:time-series-methods}.


\begin{table}[t]
\centering
\def\arraystretch{1}
\rowcolors{2}{white}{gray!25}
\begin{tabular}{
	>{\raggedright\arraybackslash}m{4cm}
	>{\centering\arraybackslash}m{3.5cm}
	>{\centering\arraybackslash}m{3.5cm}
	>{\centering\arraybackslash}m{3.5cm}
	}
\cellcolor{gray!65}\textbf{Method} &
\cellcolor{gray!50}\textbf{Preprocessing time complexity} &
\cellcolor{gray!50}\textbf{Runtime complexity} & 
\cellcolor{gray!50}\textbf{Space complexity} 
\\
	\cellcolor{gray!40}\textbf{Constant} &
	$\mathcal{O}(1)$ &
	$\mathcal{O}(1)$ &
	$\mathcal{O}(1)$ 
	\\
	\cellcolor{gray!15}\textbf{Linear} &
	$\mathcal{O}(1)$ &
	$\mathcal{O}(w)$ &
	$\mathcal{O}(1)$ 
	\\
\cellcolor{gray!40}\textbf{Simple Mean (SM)} &
	$\mathcal{O}(h)$ &
	$\mathcal{O}(1)$ & 
	$\mathcal{O}(1)$ 
	\\
\cellcolor{gray!15}\textbf{Exponential Smoothing (ES)} &
	$\mathcal{O}(k^3~h)$ &
	$\mathcal{O}(w)$ & 
	$\mathcal{O}(1)$ 
	\\
\cellcolor{gray!40}\textbf{ARIMA$(p, d, q)$} &
	$\mathcal{O}(k^3~h^2)$ &
	$\mathcal{O}((p + q)~w)$ & 
	$\mathcal{O}(max(p, q + 1))$ 
\end{tabular}
\caption{List of forecasting methods and their complexities}
\label{table:time-series-methods}
\end{table}

\subsection{Constant predictions}

As explained before, the conventional assumption that the measured values did 
not change on the absence of data is nothing else than an application of the 
\Constant{} method.
In other words, the \Constant{} method is widely (and inadvertently) adopted in 
sensor networks due to the low bandwidth links and the occasional low 
computational capabilities of the sensor networks, because it is not necessary 
to set up any model nor calculate parameters in order to assume that the 
measured value is simply the same as the last one received.
Given its practicality and low complexity, this is the most common method 
adopted in sensor networks, became the default method and has been rarely 
challenged in recent works.
The common sense, however, does not measure its limitations, weaknesses or 
the restrictions of such an option.
For example, we observed that in $46.05\%$ of the cases (i.e., in $35$ out of 
$76$ scenarios), other prediction models showed a statistically significant 
positive difference (with $95\%$ confidence intervals) when compared with the 
\Constant{}.
Therefore, we decided to adopt the \Constant{} method as a baseline and
explicitly compare its results with the other methods later.

\input{accuracy-constant-mape.aux}

As shown in Table~\ref{table:constant-accuracy-mape}, the MAPEs of the 
\Constant{} predictions vary according to the data set.
The MAPE is calculated as $\frac{1}{n} \sum_{t=1}^{n} \left| 100 
e_t / x_t \right|$; hence smaller values represent more accurate forecasts.
As an example, the forecast of $1$ value has very low relative error in the 
\Intel{} dataset ($0.00965\%$), and it is near $1\%$ when the window reaches 
its maximum value, i.e., around $1000$ times larger.
On the other hand, the \Ball{} dataset presents larger MAPE when there is only 
one value to forecast ($19.9\%$), but it increases slower when the window 
increases, e.g., it is ``only'' $17.5$ times larger ($347.395\%$) in the 
scenario with $1000$ values.

Regardless of the network types and the data sources, the explanation for such 
variances lies in the composition of the data.
Our results corroborate the initial intuition after observing the data from the 
datasets in Figure~\ref{fig:datasets}: the \Constant{} method fails 
more often in the \Ball{} than in the other scenarios, since the former has 
sharper curves where the values rapidly increase and decrease often.

\afterpage{
\begin{figure}[t]
	\centering
	\begin{subfigure}[t]{\textwidth}
		\centering
		
\includegraphics[width=\textwidth]
{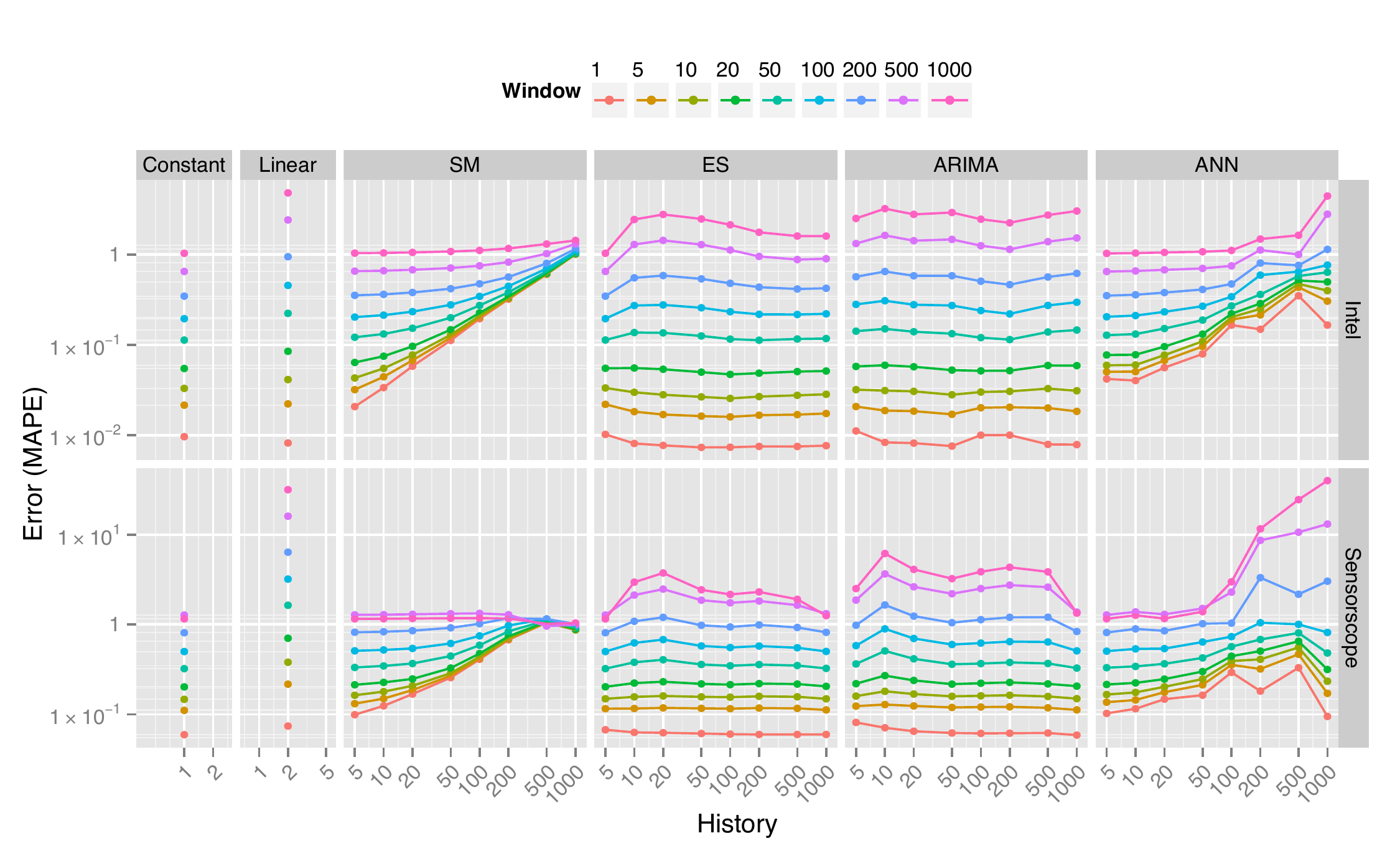}
		\caption{Accuracy in the monitoring applications.}
		\label{fig:plot-intel-vs-methods}
	\end{subfigure}%
	\qquad	
	\begin{subfigure}[t]{\textwidth}
\includegraphics[width=\textwidth]
{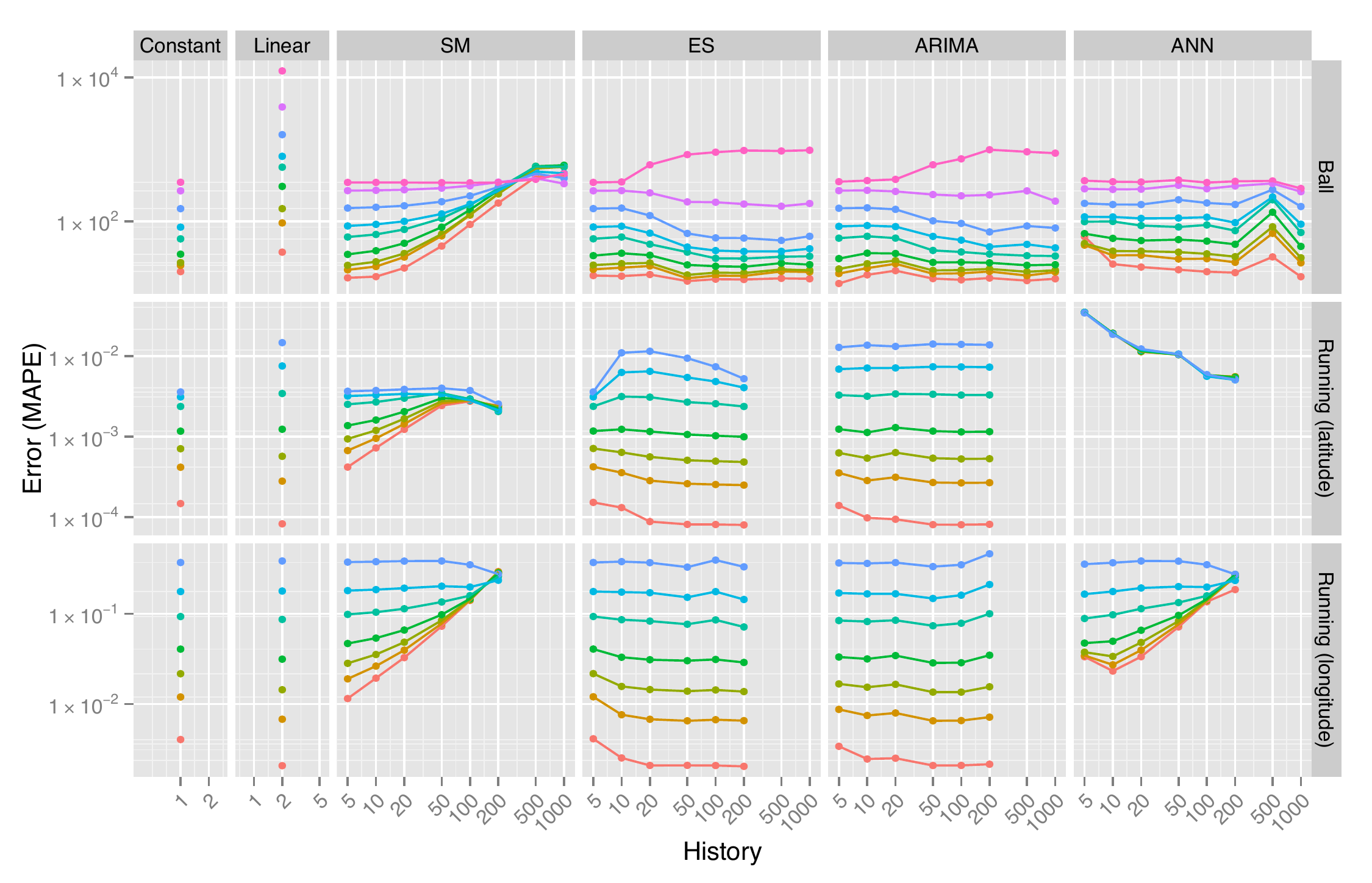}
	\caption{Accuracy in the tracking applications.}
	\label{fig:plot-running-longitude-vs-methods}
	\end{subfigure}
	\caption{Comparison between the \Constant{} and the methods that 
forecast in polynomial time.}
	\label{fig:constant-vs-all}
\end{figure}

\clearpage
}
%
%
%

%
%


\subsection{Constant predictions vs. Linear and Simple Mean}

As the \Constant{} method uses only the last value and the \LinearMethod{} 
method uses only the last two values to forecast, we did not include the 
complete history lengths in Figure~\ref{fig:constant-vs-all}.
In general, the \Constant{} method was more accurate than the \LinearMethod{} 
method.
The greatest difference between them was observed in the \Ball{} dataset, when 
the MAPE is more than one order of magnitude greater after adopting the 
\LinearMethod{} method, due to the noisy values.
The exceptions were in the \Intel{} dataset when the window had length 
equal to $1$ and in the \Running{} dataset when the window length was small 
(between $1$ and $10$).
More specifically, there is only one statistically significant improvement 
(with $95\%$ confidence intervals) when the \LinearMethod{} method is adopted in 
the \Running{ (latitude)} dataset when the window length is equal to $1$.
In this case, the MAPE is $8.255\cdot10^{-5}$ while it is $14.74\cdot10^{-5}$ 
when the \Constant{} method was adopted.

The SM method mainly forecasts as accurately as the \Constant{} in the \Ball{} 
dataset with small history length (between $5$ and $20$), when the data 
usually has the stationarity property.
In fact, the SM was the best option only in $7.33\%$ of the scenarios and the 
greatest improvement was in the \Running{ (latitude)} dataset with history 
length equal to $200$ and window length equal to $100$, where the SM method 
could reduce by $31.46\%$ the error generated by the \Constant{} method.
In most of the cases, the SM accuracy decreases noticeably when the history 
length is increased, because the data is neither stationary nor normally 
distributed.
On a side note, sensed data is rarely stationary or normally distributed, which 
exposes the limitation of this method for general use cases.

\subsection{Constant predictions vs. Exponential Smoothing vs. ARIMA}

In comparison with the ES, the \Constant{} method forecasts less accurately 
when the window is smaller, i.e., the window length is between $1$ and $10$ 
(in exceptional cases, $20$).
Moreover, we did not observe significant improvements in the accuracy of the ES 
method when the history grew larger than $50$, i.e., there was no advantage in 
using $100$ or $200$ values instead of only the last $50$.

In comparison with the ARIMA, the \Constant{} method was significantly 
outperformed most of the times when the window length has been set to $1$ 
or $5$.
The only exceptions were when the history length was $100$ or $200$, even 
though the accuracy was very similar (less than $0.0004 \%$ of relative
difference).
In addition, the ARIMA method had better accuracy when the history was at least 
$10$ times longer than the window.
For instance, when the history was at least $10$ times larger than the window, 
the average MAPE was $0.395\%$; and $1.732\%$ otherwise.

There were only few specific cases when the ES and the ARIMA models clearly 
outperformed the \Constant{} method, i.e., there was a statistically 
significant difference between their results with $95\%$ confidence intervals:
in the \Ball{} dataset, when $50$ and $100$ observations were used to predict 
the next $20$ measurements; and in the \Running{ (longitude)} dataset, when the 
next $20$ values were forecast.
Considering the different datasets, the \Constant{} method was always the 
best option in the \Sensorscope{} and in $4$ out of $20$ scenarios of the 
\Intel{} data, which can be explained by the nature of the data, i.e., in such 
cases the sensors were programmed to transmit every time interval, regardless 
the changes in their measurements, resulting on several consecutive 
similar measurements.
On the other hand, the forecasts accuracy over the \Ball{} data were 
improved in all the scenarios in which we used more than $10$ measurements and 
predicted less than $120$ values (i.e., $60\%$ of the cases).
Furthermore, in the \Running{} datasets, it was always possible to find 
models that forecast more accurately than the \Constant{} method.

As a counterpoint to the improvements explained above, both ES and ARIMA had 
lower accuracy when the window was larger than $100$ (either $500$ or $1000$), 
except in the \Ball{} database, in which they still provided accurate forecasts 
when the window length was $500$.
In conclusion, none of these methods always outperformed the others, but ES and 
ARIMA were regularly more accurate than the \Constant{} for small window 
lengths (i.e., $1$, $5$ and $10$), regardless of the scenario and the 
application.

\subsection{Constant predictions vs. Artificial Neural Networks}

As explained before, ANNs are soft computing solutions and have no upper limit 
time to compute their forecasts.
The lack of guarantee in terms of time required to converge into a final model 
can be also observed in the accuracy, as shown in 
Figure~\ref{fig:constant-vs-all}.
In general, there are few cases in which the ANN is more accurate than the 
\Constant{}, and most of them happen when the history has at least $100$ 
values, such as in the \Ball{} dataset with window $1$, $500$ and $1000$ and 
history $1000$.
Additionally, in the \Running{ (longitude)} dataset, \Constant{} is 
outperformed when the window has length $200$.

Overall, we notice that there is a large difference in the results between the 
monitoring and the tracking applications.
In the latter case, there is a clear decreasing trend in the errors when a 
longer history is adopted.
Because the tracking data is always changing and providing new information 
to the model, the ANN requires less values to capture their trends in the short 
term.
However, even in the largest history length analyzed ($2000$ values), the 
ANNs did never provide predictions more accurate than the \Constant{} method, 
independently of the scenario under consideration.

\section{Effectiveness of the forecasts}
\label{sec:feasibility}

One contribution of this work is a study that considers the data heterogeneity 
inherent to the sensor networks, which is illustrated by the datasets 
encompassing different data types and originated in different scenarios.
However, the differences between sensor networks are not only in the data 
values, but also in their requirements.
For example, a sensor network that is measuring the temperature in a data 
center might require higher precision than one placed in the mountains, because 
the indoor temperature may be used to control an air conditioning system that 
avoids damages from excessive heat.
Hence, the comparison between the MAPEs is not enough to translate how useful 
and effective is the use of forecasting methods in sensor networks.
In other words, the MAPEs illustrate the numerical differences in the forecasts 
accuracy of the methods observed, but they do not provide the practical benefit 
that they would represent in a real scenario.

\begin{figure}[t]
	\centering
	\includegraphics[width=0.6\textwidth]{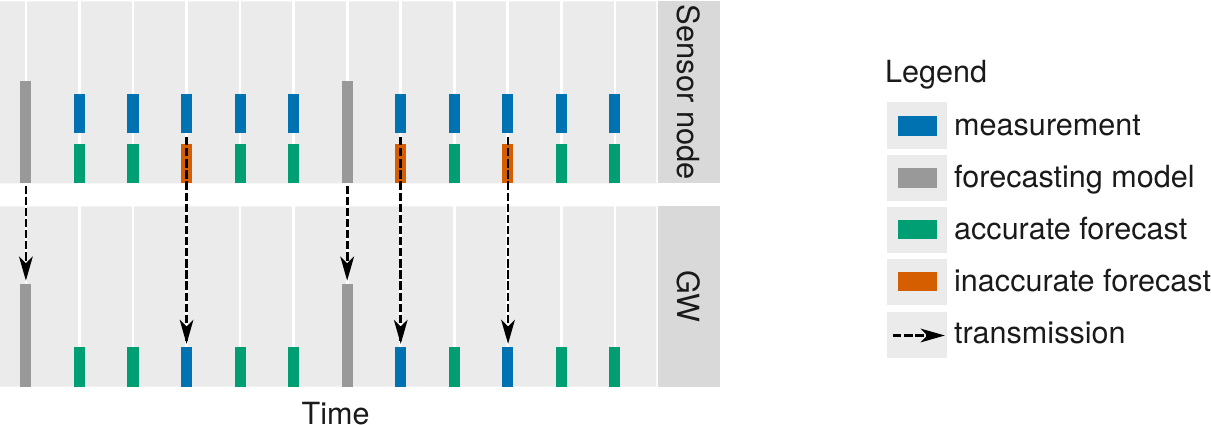}
	\caption{In a dual prediction scheme, sensor nodes transmit new forecasting
	models every time interval. A measurement is transmitted only if its
	forecast is inaccurate.}
	\label{fig:timeline}
\end{figure}

Therefore, in this Section, we evaluate the effectiveness of the forecasts and 
conclude which algorithms actually benefit the sensor networks.
We assume a dual prediction scheme~\cite{Bogliolo2014} illustrated in 
Figure~\ref{fig:timeline}. In this scheme, a sensor node has two tasks: 
\begin{inparaenum}[(i)] 
	\item make measurements; and
	\item fit a forecasting model, based on a set of measurements made beforehand.
\end{inparaenum}
Once the minimum number of measurements is collected, the sensor node is able
to fit a forecasting model and transmit it to the \Gateway{}.
Using the same forecasting model, the \Gateway{} is able to forecast the next
measurements locally without having to receive them from the sensor node.
From this point, the sensor node will only transmit a measurement if its
forecast is inaccurate, i.e., the forecast differs by more than a fixed threshold 
from the current measurement.

Our conclusions will be based on the estimated number of transmissions that 
could be saved in each case, which represents a cost-benefit relationship of 
adopting a forecasting method.
To make such an evaluation, we estimate how many transmissions could be 
avoided if the \Constant{} was used and compare it with the ES and the ARIMA 
methods, which were the only methods that could forecast as accurate as the 
former one in our previous observations.

\subsection{Acceptance threshold}

In order to define whether a transmission could be avoided, it is necessary to 
annotate, firstly, which values are relevant in a scenario.
In general, if the absolute difference between the current measurement and the 
last one transmitted is smaller than a certain $\Delta_\text{min}$ (the 
\emph{acceptance threshold}), the current measurement does not provide valuable 
information to the network and its transmission might be avoided.
Using the \emph{acceptance threshold}, it is possible to infer the potential 
reduction in the number of transmissions if we consider that a prediction is 
accurate whenever the absolute difference from the real value is smaller than 
$\Delta_\text{min}$.



%

\subsection{Results without quality loss}

\begin{table}[t]
\centering
\def\arraystretch{1}
\rowcolors{2}{white}{gray!25}
\begin{tabular}{
>{\raggedright\arraybackslash}m{3cm}
>{\centering\arraybackslash}m{1.2cm}
>{\centering\arraybackslash}m{1.2cm}
>{\centering\arraybackslash}m{1.2cm}
}
\cellcolor{gray!65}\textbf{Data set} &
\cellcolor{gray!50}\textbf{Group 1} &
\cellcolor{gray!50}\textbf{Group 2} &
\cellcolor{gray!50}\textbf{Group 3} 
\\ 
\cellcolor{gray!40} 
\textbf{\Intel{}} &  
{$ 58.13\% $} &  
{$ 46.36\% $} &  
{$ 60.14\% $} \\ 
\cellcolor{gray!15} \textbf{\Sensorscope{}} & 
{$ 61.36\%  $} &  
{$ 33.02\% $} &  
{$ 33.78\% $} \\ 
\end{tabular}
\caption{Percentage of absolute differences between consecutive measurements 
that are below the \emph{acceptance threshold} in the dataset groups used in 
the 
experiments.}
\label{table:oversampling-groups}
\end{table}

As described before, one of the goals of our study is to observe whether it is 
possible to reduce the number of transmissions using forecasting methods 
without reducing the quality of the measurements.
To simulate that, we will adopt the smallest values possible for the 
$\Delta_\text{min}$: the sensors' resolution used in each experiment.
A sensor's resolution is the smallest measurement that can be indicated 
reliably, e.g., if a temperature sensor's resolution is $0.01^{o}$C, it cannot 
precisely measure the difference between $20.001^{o}$C and $20.007^{o}$C.
In this case, we could assume that any change smaller than $0.01^{o}$C in the 
temperature is never relevant; and hence if a forecast differs by less than 
$0.01^{o}$C from the real observation it is accurate and will not trigger a 
transmission from the sensor node to the \Gateway{}.
The following values of $\Delta_\text{min}$ were considered in each dataset:

\begin{itemize}
 \item \textbf{\Intel{} dataset}: $0.01^{o}$C. 
Since the Mica2Dot's specification does no include the temperature sensor's 
resolution~\cite{MICA2DOT}, we defined the $\Delta_\text{min}$ as the minimum 
difference between any pair of measurements observed in the data.
On a side note, we highlight that it is the same resolution of Sensirion 
SHT11, a temperature sensor broadly used in several wireless sensor 
nodes similar to Mica2Dot~\cite{Crossbow2007}. 
 \item \textbf{\Sensorscope{} dataset}: $0.045^{o}$C. 
According to the TinyNode's specification~\cite{ShockfishSA2005}, it uses an 
analog temperature sensor (LM20) that can measure from $-55^{o}$C to $130^{o}$C 
and the microcontroller has a 12-bit analog-to-digital converter. 
Therefore, we calculated the sensor's resolution as $(130 - (-55)) / 
2^{12}~^{o}$C.
 \item \textbf{\Ball{} dataset}: $0.001$ meter. 
As a reference, we adopted the specification of the AR3000 sensor, a long range 
laser sensor that can accurately measure distances of up to 300 
meters~\cite{SchmittIndustries2010}.
 \item \textbf{\Running{} datasets}: $8.38 \cdot 10^{-8}$ degree.
Since the sensor's specification is not publicly available, we defined the 
$\Delta_\text{min}$ as the minimum difference between any pair of measurements 
observed in the data. 
This value represents nearly $1$ centimeter of distance along the latitude 
direction and around $0.7$ centimeters along the longitude 
orientation~\cite{Xu2007}.
\end{itemize}

We highlight that only the \Intel{} and the \Sensorscope{} datasets had 
consecutive values such that their absolute difference is smaller than the 
\emph{acceptance threshold}, due to the consecutive similar values and the 
missing values filled beforehand.
Table~\ref{table:oversampling-groups} shows the percentage of absolute 
differences between consecutive measurements that are below the 
\emph{acceptance threshold}.

\begin{figure}[t]
	\centering
	
\includegraphics[width=\textwidth]{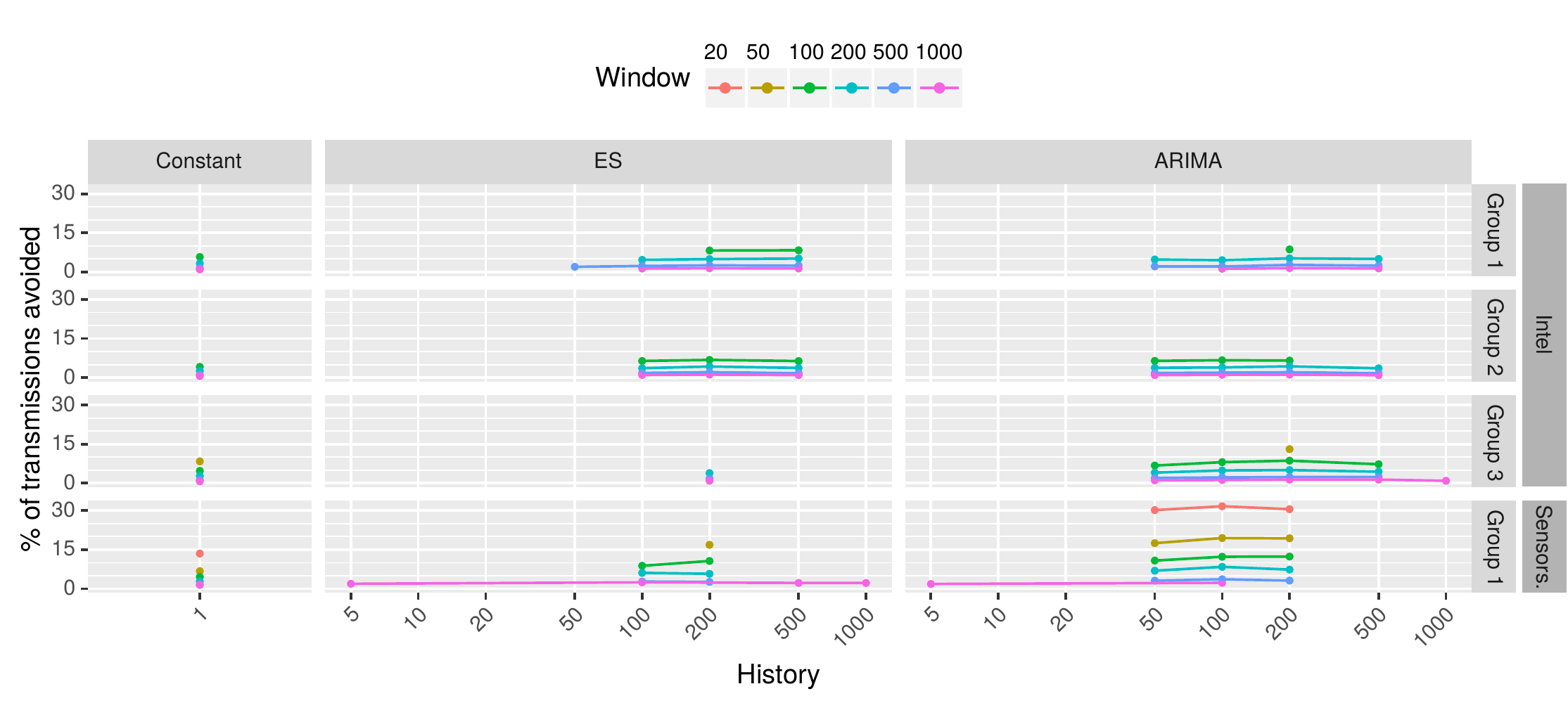}
	\caption{The percentage of transmissions that could be avoided using 
each forecasting method. We included only the cases in which ES and ARIMA could 
save at least 2 transmissions more than the \Constant{} method.}
	\label{fig:constant-es-arima-auto-aicc-tx}
\end{figure}

Differently from the \Constant{} method, ES and ARIMA are based on models and 
use sets of parameters to forecast new values.
In a real sensor network where the forecasting model is generated in the sensor 
nodes, it means that the same parameters must be shared with the \Gateway{}.
That is, the parameters must be transmitted; and this can either be done 
by triggering a new transmission or simply by attaching them to a 
measurement transmission.
Hence, using ES or ARIMA with window length equal to $1$ cannot reduce the 
number of transmissions in the sensor nodes, given that the model would have to 
be updated after each measurement.
In fact, updating the model after each measurement would eventually increase 
the energy consumption in the sensor nodes, given that their parameters require 
an extra computational time inexistent in the \Constant{} method.

Furthermore, depending on the scenario, an inaccurate forecast may trigger the 
transmission of the real measurement to the \Gateway{}. 
Then, if the \Constant{} method has been adopted, the new measurement can be 
used to forecast new values, reseting the forecasting window.
On the other hand, the ES and the ARIMA methods may trigger (at least) 
one extra transmission to establish the same forecasting model in the sensor 
node and in the \Gateway{}.
Given the occasional extra transmission, we make a fair comparison by focusing 
only on the scenarios in which the ES and the ARIMA methods could reduce at 
least $2$ transmissions more than the \Constant{} method.

Finally, considering the window length as the maximum number of measurements 
that can be transmitted (in case that all forecasts are inaccurate), 
Figure~\ref{fig:constant-es-arima-auto-aicc-tx} shows the percentage of 
transmissions that could be avoided using the \Constant{}, ES and ARIMA 
methods.
For this representation, we did not consider any extra transmission that could 
be occasionally required to transmit the forecasting models' parameters.
As explained above, to improve the readability, we included only the cases in 
which the ES and the ARIMA methods accurately predicted, on average, at least 
$2$ values more than the \Constant{} method.

The results show that the \Constant{} method is the best option when the window 
length is smaller than $20$.
Curiously, as discussed in the previous Section, the \Constant{} method was 
regularly less accurate than the ES and the ARIMA methods when the window 
length was $1$, $5$ or $10$.
This illustrates the importance of analyzing the real effectiveness of the 
forecasts when applied to real use cases.

Considering all scenarios, the greatest improvement of a forecasting method in 
comparison with the \Constant{} method was observed in \Sensorscope{} when the 
window length was set to $20$: the number of transmissions avoided using ARIMA 
was $18.1\%$ greater than using \Constant{} method, i.e., the ARIMA model could 
accurately forecast nearly $3.9$ transmissions more than the \Constant{} method.
In this scenario, the ARIMA method could be used to avoid an average of $6.33$ 
transmissions every $20$ measurements, which represents a reduction of $31.65\%$ 
in the number of transmissions.

In terms of history length, we observed that increasing the number of values in 
the history 
from $100$ to $200$ increased the number of avoided transmissions in $70\%$ of 
the cases illustrated in the plot.
In general, ES could reduce more transmissions when the history length 
was between $100$ and $200$, while ARIMA also performed well using only $50$ 
values.
These results, together with the observations obtained in the previous tests, 
suggest that both forecasting methods can be more accurate and bring improvements 
when the history length is increased up to $200$ (and not to $500$ or $1000$).

In conclusion, our experimental results show that it is possible to reduce 
the number of transmissions using forecasting methods without losing any quality 
of the measurements provided by the sensors.
The effectiveness of such methods, however, may vary from sensor to sensor even 
if the environment and the phenomena observed is the same, for example, in the 
\Sensorscope{} dataset only the sensor node that sensed the \emph{Group 1} data 
could have its transmissions reduced.
The decision about whether to adopt forecasting methods or not can be made based
on the methodology presented in this paper, after observing the sensor nodes' 
computing capacities and the environment under observation.

Finally, we observed that, as shown in Table~\ref{table:oversampling-groups}, 
in the four cases illustrated in the plot, more than $45\%$ of absolute differences 
between consecutive measurements were below the \emph{acceptance threshold}.
Such a coincidence suggests that the forecasting methods can be more effective 
under some circumstances.

\subsection{Results with different sensors' resolution}

\begin{table}[t]
\centering
\def\arraystretch{1}
\rowcolors{2}{white}{gray!25}
\begin{tabular}{
>{\raggedright\arraybackslash}m{3cm}
>{\centering\arraybackslash}m{3.84cm}
>{\centering\arraybackslash}m{3.84cm}
>{\centering\arraybackslash}m{3.84cm}
}
\cellcolor{gray!65}\textbf{Data set} &
\cellcolor{gray!50}\textbf{Group 1} &
\cellcolor{gray!50}\textbf{Group 2} &
\cellcolor{gray!50}\textbf{Group 3} 
\\ 
\cellcolor{gray!40} 
\textbf{\Intel{}} &  
{$ 0.0098^{o}$C} &
{$ 0.0101953^{o}$C} &
{$ 0.009653105^{o}$C} \\
\cellcolor{gray!15} \textbf{\Sensorscope{}} & 
{$ 0.02137582^{o}$C} &
{$ 0.0840074^{o}$C} &
{$ 0.07867205^{o}$C} \\
\cellcolor{gray!40} \textbf{\Ball{}} & 
{$ 0.9410768 $ meter} &
{$ 0.9699227 $ meter} &
{$ 0.9555685 $ meter}\\
\cellcolor{gray!15} \textbf{\Running{ (latitude)}} & 
$ 3.35276 \cdot 10^{-5} $ degree ($\sim3.72$ meters)&
$ 3.704801 \cdot 10^{-5} $ degree ($\sim4.11$ meters)&
$ 3.43658 \cdot 10^{-5} $ degree ($\sim3.81$ meters)\\
\cellcolor{gray!40} \textbf{\Running{ (longitude)}}& 
$ 0.0001015048 $ degree ($\sim8.49$ meters)&  
$ 0.0001128204 $ degree ($\sim9.43$ meters)&  
$ 0.0001072884 $ degree ($\sim8.97$ meters)\\ 
\end{tabular}
\caption{New sensors' resolution and \emph{acceptance thresholds} set in order 
to have similar consecutive values in $50\%$ of the time.}
\label{table:new-sensors-resolutions}
\end{table}
\vspace{0.1cm}

The results illustrated above suggest that if the sensors sampled more often 
and measured more ``similar'' values, the ES and the ARIMA methods would be 
able to reduce more transmissions than the \Constant{} method.
In order to observe the impact of the similarity between consecutive data samples, 
we simulated a change in the sensors' 
resolution in each \emph{Group} to values such that the chances of observing 
two similar consecutive values was $50\%$.
For example, the sequence of temperature values 
$\{20.1^{o}\text{C}$, 
$20.1^{o}\text{C}$, 
$20.4^{o}\text{C}$, 
$20.6^{o}\text{C}$, 
$21.5^{o}\text{C}$, 
$21.6^{o}\text{C}$, 
$21.8^{o}\text{C}\}$ 
measured by a sensor with resolution equal to $0.1^{o}\text{C}$ has only one 
pair of similar consecutive values (out of $6$).
However, if the sensor's resolution was $0.5^{o}\text{C}$, the same sequence 
would be measured as 
$\{20.0^{o}\text{C}$, 
$20.0^{o}\text{C}$, 
$20.5^{o}\text{C}$, 
$20.5^{o}\text{C}$, 
$21.5^{o}\text{C}$, 
$21.5^{o}\text{C}$, 
$22.0^{o}\text{C}\}$
where half of the values are the same as the last one observed.
On the one hand it represents a change in the \emph{acceptance threshold}, 
but on the other hand it is similar to changing the sensors' sampling rate to 
a value in which the measurements are the same as the last one measured in 
$50\%$ of the time.

\begin{table}[t]
\centering
\def\arraystretch{1}
\rowcolors{2}{white}{gray!25}
\begin{tabular}{
>{\raggedright\arraybackslash}m{3cm}
>{\centering\arraybackslash}m{1.5cm}
>{\centering\arraybackslash}m{1.5cm}
>{\centering\arraybackslash}m{4.25cm}
>{\centering\arraybackslash}m{4.25cm}
}
\cellcolor{gray!65}\textbf{Data set} &
\cellcolor{gray!50}\textbf{Window length} &
\cellcolor{gray!50}\textbf{History length} &
\cellcolor{gray!50}\textbf{Saved transmissions using the \Constant{} method} &
\cellcolor{gray!50}\textbf{Saved transmissions using another forecasting 
method} 
\\ 
\cellcolor{gray!40} 
\textbf{\Intel{}} &  
{$20$} &
{$200$} &
{$8.59\%$} &
{$20.91\%$ using ARIMA} \\
\cellcolor{gray!15} \textbf{\Sensorscope{}} & 
{$20$} &
{$100$} &
{$6.87\%$} &
{$21.08\%$ using ARIMA} \\
\cellcolor{gray!40} \textbf{\Ball{}} & 
{$50$} &
{$1000$} &
{$14.95\%$} &
{$43.03\%$ using ARIMA} \\
\cellcolor{gray!15} \textbf{\Running{ (latitude)}} & 
{$50$} &
{$100$} &
{$0.58\%$} &
{$5.21\%$ using ES} \\
\cellcolor{gray!40} \textbf{\Running{ (longitude)}}& 
{$5$} &
{$200$} &
{$13.91\%$} &
{$51.74\%$ using ARIMA} \\
\end{tabular}
\caption{Highest improvements in the percentage of saved transmissions.}
\label{table:highest-savings-new-sensors-resolutions}
\end{table}

After adjusting the sensors' resolution to the values observed in 
Table~\ref{table:new-sensors-resolutions}, we obtained the desired scenario 
described above and the results shown in 
Figure~\ref{fig:q50-constant-es-arima-auto-aicc-tx}.
As well as before, to improve the readability, we included only the cases in 
which the ES and the ARIMA methods accurately predicted, on average, at least 
$2$ values more than then \Constant{} method.

We can observe that it is possible to reduce the number of transmissions 
regardless of the application type, even though each dataset has a different 
average percentage of avoided transmissions per window length.
Table~\ref{table:highest-savings-new-sensors-resolutions} shows the greatest 
reductions in the number of transmissions after setting new resolutions for the 
sensors. 
In the best case, the ARIMA method can reduce more than $50\%$ of the 
transmissions in the \Running{ (longitude)} dataset, which represents a 
significant improvement especially taking into consideration its potential to 
improve the end-to-end throughput and reduce the delays in other tracking 
applications.

\afterpage{
\begin{figure}[t]
	\centering
	
\includegraphics[width=\textwidth]{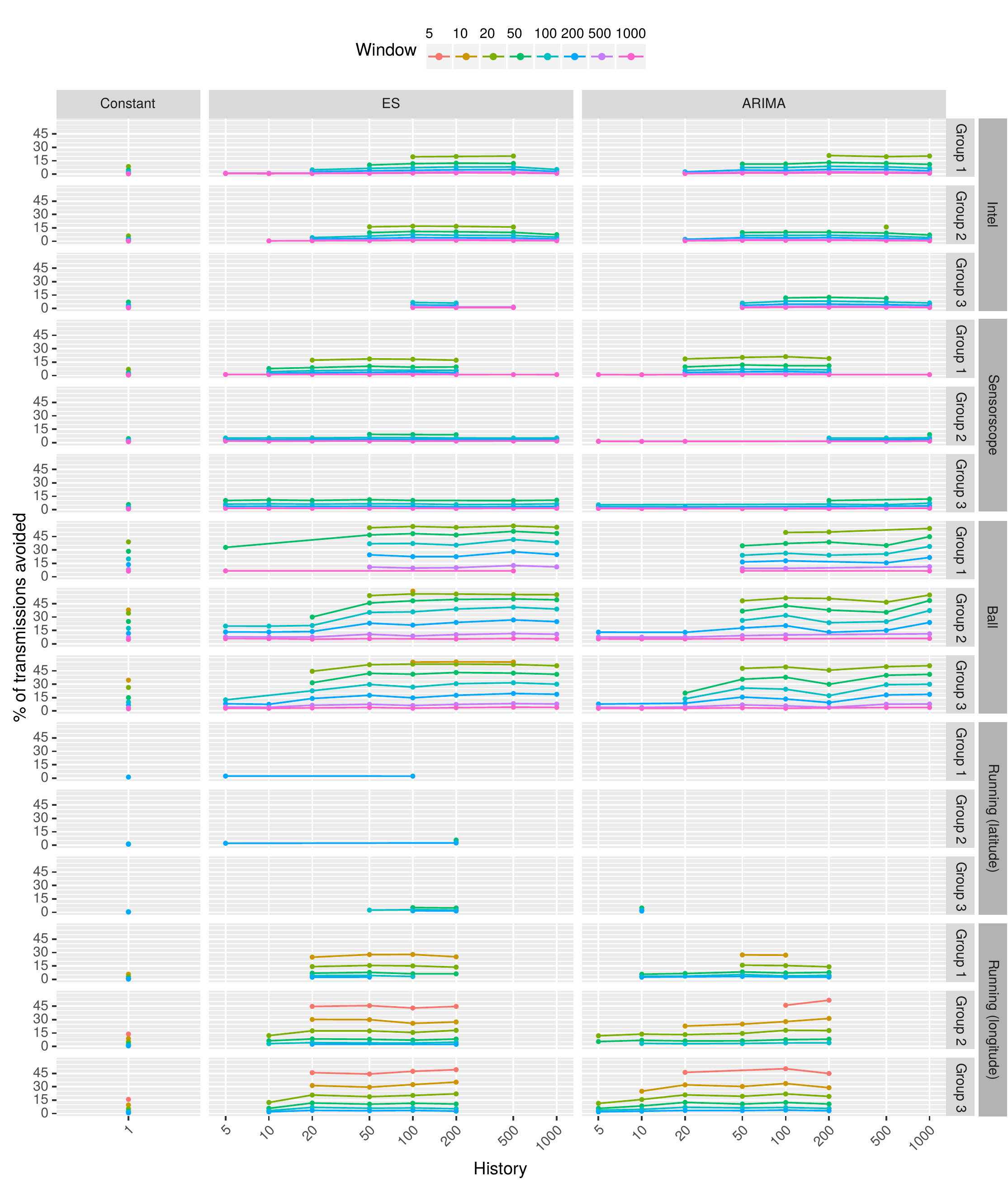}
	\caption{In cases where the chances of measuring two consecutive values 
is $50\%$, it is possible to have a high reduction in the number of 
transmissions without reducing the quality of the measurements provided by the 
sensor network.}
	\label{fig:q50-constant-es-arima-auto-aicc-tx}
\end{figure}
\clearpage
}

Furthermore, in comparison with the previous results, now it is possible to 
observe cases in which there is a reduction in the number of transmissions with 
smaller window lengths ($5$ and $10$).
Probably due to the nature of the measurements and an eventual seasonality, 
when the window length was set to $1000$, the ES and the ARIMA methods could 
reduce the number of transmissions only in the monitoring applications.
Again, the great majority of the improvements using ES and ARIMA needed between 
$50$ and $200$ values in the history.

\section{Conclusion}
\label{sec:conclusion}

In this work, we provided a broad study about the adoption of forecasting 
methods in sensor networks.
First, we presented the importance of reducing the number of transmissions in 
sensor networks and assessed the forecasting as a potential candidate to assume 
such a responsibility.
Related works in the field that adopted forecasting methods to reduce the data 
transmission in the sensor networks corroborated our initial guess and motivated 
our further experiments.
In order to explore the diversity of the sensor network and IoT applications, 
we based our conclusions on the experimental results over $4$ datasets from 
different types of origins, embracing the data heterogeneity expected from real 
implementations, such as various data types represented in different units of 
measurement and collected from different applications.
Finally, after observing the relative accuracy and the real efficacy of the 
different forecasting methods, we concluded that adopting complex forecasting 
methods in the sensors can be as promising as we initially suspected: using 
forecasting methods, it is possible to reduce the number of transmissions 
without reducing the quality of the measurements provided by the sensor 
networks.

As discussed before, the computing times of ES and ARIMA are respectively 
proportional to the length and the squared length of the set of values used to 
generate the forecasting models.
Hence, the reduction in the number of transmissions can be especially achieved when 
the sensor nodes have high computational power to compute complex algorithms,
since it was necessary between $50$ and $200$ values to generate forecasting 
models that could effectively reduce the number of transmissions.
In other words, the simplest wireless sensor nodes, such as the 
TelosB~\cite{XBowTelosB}, may not have enough computing power to process these 
forecasting methods or enough memory to store the necessary values needed to 
reduce the number of transmissions to the \Gateway{}.
We conclude that the evolution of the sensor nodes will make it possible to 
forecast their data and has potential to support the exponential growth of IoT, 
regardless of their limited access to information that could be provided by 
neighboring sensor nodes or external data sources.
In other words, the number of transmissions can be reduced by simply exploiting 
the sensors' proximity to the origin of the data and their computational power 
without losing the quality of the measurements, because the current 
state-of-the-art forecasting methods are accurate enough to substitute the real 
measurements made by the sensors.

\section*{Acknowledgment}

This work has been partially supported by the Spanish Government through the 
project TEC2012-32354 (Plan Nacional I+D), by the Catalan Government 
through the project SGR-2014-1173 and by the European Union through the 
project FP7-SME-2013-605073-ENTOMATIC.

\bibliographystyle{IEEEtran}
\bibliography{IEEEabrv,bibliography}

\end{document}